\documentclass[useAMS,usenatbib,usegraphicx]{mn2e}

\usepackage{rotating}

\def\hd{HD\,93161}
\def\l{$\lambda$}
\def\ha{H$\alpha$}
\def\he{He\,{\sc i}}
\def\hee{He\,{\sc ii}}
\def\n{N\,{\sc{iii}}}
\def\si{Si\,{\sc{iv}}}
\def\s{S\,{\sc{iv}}}
\def\ciii{C\,{\sc{iii}}}
\def\civ{C\,{\sc{iv}}}
\def\o{O\,{\sc{iii}}}
\def\kms{km\,s$^{-1}$}

\def\xmm{{\sc XMM}\emph{-Newton}}
\def\ros{\emph{{\sc ROSAT}}}

\title[The spectroscopic binary \hd A and its visual companion \hd B]{Optical spectroscopy of X-Mega targets in the Carina Nebula - V. The spectroscopic binary \hd A and its visual companion \hd B \thanks{Based on observations collected at the European Southern Observatory (La Silla, Chile), at the Cerro Tololo Inter-American Observatory (CTIO), and with \xmm, an ESA Science Mission with instruments and contributions directly funded by ESA Member States and the USA (NASA).}}

\author[Y. Naz\'e, I.I. Antokhin, H. Sana, E. Gosset, and G. Rauw]{
Y. Naz\'e$^{1}$\thanks{E-mail: naze@astro.ulg.ac.be},
I.I. Antokhin$^{2}$,
H. Sana$^{1}$\thanks{Research Fellow FNRS (Belgium)},
E. Gosset$^{1}$\thanks{Research Associate FNRS (Belgium)}\thanks{Visiting Astronomer, CTIO, National Optical Astronomy Observatories (NOAO). NOAO is operated by the Association of Universities for Research in Astronomy, Inc.\ under contract with the National Science Foundation.},
and G. Rauw$^{1}$\footnotemark[4]\footnotemark[5] \\
$^{1}$Institut d'Astrophysique et de G\'eophysique, Universit\'e de Li\`ege, B\^at. B5c,  All\'ee du VI Ao\^ut 17, B-4000 Li\`ege, Belgium.\\
$^{2}$Sternberg Astronomical Institute, Moscow University, Universitetskij
Prospect 13, Moscow 119899, Russia.}

\begin{document}


\pagerange{\pageref{firstpage}--\pageref{lastpage}} \pubyear{2005}

\maketitle

\label{firstpage}

\begin{abstract}
We present the analysis of an extensive set of high-resolution spectroscopic observations of \hd, a visual binary with a separation of 2\arcsec.
\hd A is a spectroscopic binary, with both components clearly detected throughout the orbit. The primary star is most probably of spectral type O8V, while the secondary is likely an O9V. We obtain the first orbital solution for this system, characterized by a period of 8.566$\pm$0.004~days. The minimum masses of the primary and secondary stars are 22.2$\pm$0.6 M$_{\odot}$ and 17.0$\pm$0.4 M$_{\odot}$ respectively. These values are quite large, suggesting a high inclination of the orbit. 
The second object, \hd B, displays an O6.5V((f)) spectral type and is thus slightly hotter than its neighbour. This star is at first sight single but presents radial velocity variations.\\
We finally study \hd\ in the X-ray domain. No significant variability is detected. The X-ray spectrum is well described by a 2T model with k$T_1\sim0.3$~keV and k$T_2\sim 0.7$~keV. The X-ray luminosity is rather moderate, without any large emission excess imputable to a wind interaction. 
\end{abstract}


\section{Introduction}

A few years ago began an extensive campaign to study the Carina Nebula, one of the largest concentrations of young and massive stars known in our Galaxy. This campaign was initiated by the X-Mega group \citep{COR99} in order to gain a good knowledge of the cluster's binary population prior to its observation in the X-ray domain by the \ros\ satellite. Such multiwavelength observations of spectroscopic binaries in open stellar clusters enable to constrain the fundamental parameters of the early-type stars and to investigate the interactions between their components in more depth. \\

The efforts were focused more specifically on the massive stars of the Trumpler\,16 (Tr\,16) open cluster \citep[ hereafter Papers I-IV]{ALB01,MOR01,RAU01,ALB02}. This cluster constitutes one of the best places to complete our objective, since it harbours a large number of binaries: \citet{LEV91} found at least 5 spectroscopic binaries and several radial velocity variables in Tr\,16. The (very) preliminary results of \citet{LEV91}, relying on poor quality and scarce data, were subsequently revised in the first papers of this series.\\

\begin{table*}
\centering
\begin{minipage}{140mm}
\caption{Journal of the spectroscopic observations of \hd\ A and B. Column 1 gives the instrument used and Column 2 lists the Heliocentric Julian Date (HJD) at mid-exposure. The next three columns present the phases (calculated from the \he\,\l\,4471 orbital solution with $e=0$) and the \he\,\l\,4471 RVs of the primary and secondary components. The last three columns are similar, but refer to the averaged velocities (see Section 5).
\label{tab: obs} }
\begin{tabular}{l r r r r r r r}
\hline
Instrument &  HJD$-$2\,450\,000 & $\phi_{4471}$ & RV$_1$ & RV$_2$ & $\phi_{\rm avg}$ & $\overline{\rm RV}_1$ & $\overline{\rm RV}_2$ \\
 & (days) & & \multicolumn{2}{c}{(\kms)} & & \multicolumn{2}{c}{(\kms)} \\
\hline
\multicolumn{8}{c}{\hd A}\\
\hline
B\&C  &  534.752 &  0.085  &    90.0  &$-$105.4   &    \\ 
      &  535.758$^a$\hspace*{-1.5mm}  &         &          &           &    \\
      &  536.748 &  0.318  &   111.8  &$-$167.0   &    \\
      &  537.741 &  0.434  &    81.8  & $-$83.9   &    \\
      &  538.733 &  0.550  & $-$20.1  & $-$20.1   &    \\
      &  539.661 &  0.658  &$-$151.0  &   100.9   &    \\
BME   & 1328.625 &  0.760  &$-$178.5  &   194.6   &   0.761  & $-$156.9  &   173.5   \\
      & 1329.556 &  0.869  &$-$112.2  &   127.7   &   0.870  & $-$117.3  &   125.8   \\
      & 1330.590 &  0.990  &  $-$6.7  &  $-$6.7   &   0.991  &   $-$5.5  &  $-$4.4   \\
      & 1331.674 &  0.116  &    81.1  &$-$141.9   &   0.117  &     91.3  &$-$130.2   \\
      & 1332.560 &  0.220  &   149.6  &$-$195.4   &   0.221  &    153.3  &$-$192.6   \\
FEROS & 1299.554 &  0.367  &   102.8  &$-$144.2   &   0.368  &    111.1  &$-$150.4   \\
      & 1300.563 &  0.484  &  $-$4.0  &  $-$4.0   &   0.485  &   $-$5.9  &  $-$5.0   \\
      & 1301.670 &  0.614  &$-$100.9  &   131.6   &   0.615  & $-$102.1  &   130.8   \\
      & 1302.662 &  0.729  &$-$157.6  &   204.6   &   0.731  & $-$150.3  &   198.2   \\
      & 1304.560 &  0.951  & $-$54.7  &    66.6   &   0.952  &  $-$44.7  &    62.5   \\
      & 1327.531 &  0.633  & $-$98.9  &   155.1   &   0.634  & $-$102.6  &   151.9   \\
      & 1669.577 &  0.562  & $-$55.6  &    81.9   &   0.563  &  $-$58.4  &    79.8   \\
      & 1670.539 &  0.675  &$-$139.7  &   184.2   &   0.675  & $-$143.3  &   184.2   \\
      & 1671.535 &  0.791  &$-$151.8  &   195.7   &   0.791  & $-$149.2  &   192.4   \\
      & 1672.527 &  0.907  & $-$92.7  &   121.0   &   0.907  &  $-$95.7  &   109.7   \\
      & 1673.526 &  0.023  &  $-$3.6  &  $-$3.6   &   0.024  &   $-$2.0  &  $-$1.1   \\
      & 2037.627 &  0.528  & $-$70.7  &    19.8   &   0.528  &   $-$7.9  &    18.6   \\
      & 2038.556 &  0.636  &$-$126.1  &   107.5   &   0.636  & $-$120.5  &   112.4   \\
      & 2039.628 &  0.761  &$-$159.8  &   211.4   &   0.761  & $-$156.5  &   205.7   \\
      & 2040.636 &  0.879  & $-$79.0  &   133.4   &   0.879  &  $-$93.0  &   133.5   \\
      & 2335.615 &  0.314  &   112.6  &$-$192.8   &   0.314  &    118.4  &$-$191.0   \\
      & 2336.617 &  0.431  &    24.7  &$-$102.4   &   0.431  &     52.7  & $-$72.6   \\
      & 2337.606 &  0.547  & $-$70.3  &    80.9   &   0.546  &  $-$64.7  &    49.5   \\
      & 2338.604 &  0.663  &$-$111.7  &   173.4   &   0.663  & $-$114.5  &   167.5   \\
      & 2381.573 &  0.679  &$-$138.5  &   185.4   &   0.679  & $-$139.4  &   182.4   \\
      & 2382.552 &  0.794  &$-$156.8  &   195.8   &   0.793  & $-$154.3  &   190.8   \\
      & 2383.556 &  0.911  & $-$75.7  &   105.2   &   0.910  &  $-$79.1  &   103.4   \\
      & 2782.527 &  0.486  &  $-$2.2  &  $-$2.2   &   0.485  &   $-$2.4  &  $-$0.9   \\
      & 2783.516 &  0.601  & $-$84.2  &   121.5   &   0.600  &  $-$87.9  &   121.0   \\
      & 2784.511 &  0.717  &$-$157.4  &   205.3   &   0.716  & $-$154.4  &   200.8   \\
      & 3130.515 &  0.109  &   101.2  &$-$118.2   &   0.108  &    100.7  &$-$122.4   \\
      & 3131.495 &  0.224  &   161.2  &$-$192.8   &   0.222  &    160.8  &$-$193.3   \\
      & 3132.494 &  0.340  &   141.9  &$-$170.4   &   0.339  &    141.9  &$-$170.8   \\
      & 3133.575 &  0.466  &    39.9  & $-$55.3   &   0.465  &     39.7  & $-$50.0   \\
      & 3134.513 &  0.576  & $-$65.4  &    97.7   &   0.574  &  $-$68.1  &    94.4   \\
      & 3135.544 &  0.696  &$-$146.9  &   195.2   &   0.695  & $-$147.1  &   193.1   \\
\hline
\multicolumn{8}{c}{\hd B}\\
\hline
BME   & 1330.640& & $-$4.5\\
FEROS & 2382.563& & $-$57.8\\
      & 3132.506& & $-$14.0\\
\hline
\end{tabular}\\
a. No measurement is available for \he\,\l\,4471, but the \hee\,\l\,4686 line shows RVs of 109.2 and $-$161.2~\kms\ for the primary and secondary, respectively.\\
\end{minipage}
\end{table*}

In the present paper, we analyse an extensive set of spectra of one of the still poorly studied systems of Tr\,16, \hd\ (= Tr\,16-176, CPD$-58^{\circ}$2631)\footnote{The membership of this system to Tr\,14, Tr\,16 or Collinder\,232 has been subject to controversy throughout the literature. Here we follow \citet{WAL73} in considering that it belongs to Tr\,16.}. The compound formed by \hd\ and HD\,93160 actually harbours five stellar components, noted A--E, some of which were found by speckle observations \citep{MAS98}: HD\,93160 corresponds to the C component, located at 13\arcsec\ from the AB pair; component D is at 3\arcsec\ from HD\,93160 and component E is at 8\arcsec\ of the AB pair. The latter two stars, separated by only 2\arcsec, have been known for a long time \citep[see e.g.][]{WAL73} and are generally referred to as \hd. \citet{PEN96} and \citet{HOW97} reported radial velocity (RV) variations in the UV spectrum of \hd. Unfortunately, the small separation between A and B, together with the low resolution of IUE, have not permitted to attribute these changes confidently to one member of the pair. \citet{HOW97} even suggested that both stars could be short-period binaries.  Previously, \citet{LEV91} had provided an SB1 orbital solution for \hd, without specifying to which component this solution applied. A thorough study of the system was thus eagerly awaited.\\

This paper is organized as follows: the observations and the data reduction are presented in Sect.\,2 and the spectral features of \hd\ A and B are discussed in Sections 3 and 4.  In Sect.\,5, we derive the first full orbital solution for \hd A, whereas Sect.\,6 deals with the analysis of its X-ray emission. As a next step, we examine the evolutionary status of the stars and we finally conclude in Sect.\,8. \\

\section{Observations and data reduction}
\subsection{Optical spectroscopy}
\hd\ lies in a crowded region of the Carina Nebula, with HD\,93160 just 13\arcsec\ west of the star. As noted in the introduction, \hd\ itself was found to be a visual double system, composed of two stars with similar brightnesses separated by only 2\arcsec. As in \citet{VRA}, we define component A as being the one closest to HD\,93160. 
With such a small separation between A and B, it is difficult to study these stars individually. During some observing nights, a poor seeing and/or the low angular resolution of the pointing camera did not enable us to clearly disentangle the close pair AB. Among our 45 spectra, only 3 can securely be attributed to the sole B component. The main contributor to the remaining 42 spectra is intended to be the A component but these observations are usually contaminated, to a variable degree, by the B companion. This is marked in the spectra by the presence of a third, weak component. \\

We have observed \hd\ over a period of 7 years with various instruments. 
A first set of 6 medium resolution spectra covering the wavelength range $3850 - 4800$\,\AA\ was gathered in 1997 with the ESO 1.5\,m telescope equipped with a Boller \& Chivens (hereafter B\&C) Cassegrain spectrograph. The data were obtained with a holographic grating (2400 lines/mm, ESO grating \#32) providing a reciprocal dispersion of 32.6\,\,\AA/mm. The detector was a thinned, UV flooded Loral-Lesser CCD (ESO \#39). The slit width was set to 220\,$\mu$m corresponding to 2\arcsec\ on the sky. The spectral resolution as measured on the helium-argon calibration spectra is 1.2\,\AA, corresponding to a resolving power of 3600. Typical exposure times were of the order of 10 minutes and the average signal-to-noise ratio was about 250. Part of the B\&C spectra were affected by a fringing pattern \citep{b39} that occurred over the wavelength range $4050 - 4250$\,\AA\ in this instrument configuration. Given the variability of the fringing pattern and in order to avoid amplification of the fringes in the stellar spectra, the data were not flat-fielded. All the reductions were performed using the {\sc midas} software developed at ESO. Note that only B\&C spectra were sky subtracted.\\

Six high resolution spectra of \hd\ were taken in May-June 1999 with the Bench-Mounted Echelle spectrograph (BME) fed by the 1.5\,m Ritchey-Chr\'etien telescope at CTIO. These data covered the spectral range $3750 - 5800$\,\AA. The typical S/N ratio was $\sim$60 at 5430\,\AA\ for exposure times of one hour. The detector used was a Tek CCD with 2048 $\times$ 2048 pixels of $24\,\mu$m $\times$ $24\,\mu$m and the resolving power was 45\,000.
The BME data were reduced using the IRAF\footnote{IRAF is distributed by the National Optical Astronomy Observatories.} package, following the recommendations of the BME User's Manual. A first rectification of the extracted orders was carried out with the projector flat exposures. The spectra were then  normalized by fitting a low-order polynomial to the continuum.\\

Between 1999 and 2004, we obtained 33 additional high resolution spectra at the La Silla Observatory with the Fiber-fed Extended Range Optical Spectrograph \citep*[FEROS,][]{Kaufer}. This spectrograph was installed until October 2002 at the ESO~1.5\,m telescope, and was subsequently moved to the 2.2\,m telescope. The FEROS spectra cover the wavelength range $3750 - 9000$\,\AA\, with a resolving power of 48\,000. The detector was a 2k $\times$ 4k EEV CCD with pixel size $15\,\mu$m $\times$ $15\,\mu$m. The typical exposure time was 10--15 minutes and the S/N ratio at 5650\,\AA\, is about 125--225.
The FEROS spectra were reduced using the appropriate context of the {\sc midas} environment, together with an improved reduction pipeline \citep{SHR03}. They were normalized by fitting the continuum with a polynomial of degree ranging from 2 to 6. Most of the unavoidable fringes affecting the red part of the spectra were corrected by flat-fielding, but there are a few residuals, e.g. near 6640~\AA. \\  



\subsection{X-ray data}

The core of the Carina Nebula was observed several times by the \xmm\ observatory \citep{JAN01}. We will focus here on the five datasets acquired in 2000 and 2001. The first two datasets, centered on $\eta$ Carinae, were obtained in July 2000 during satellite revolutions \#115 and \#116. The three other observations, centered on WR\,25, were taken one year later during Revs. \#283, \#284, and \#285. A preliminary study using only the first two observations was presented by \citet{ALB03}. A thorough analysis of the field relying on the entire observational dataset has been performed by \citet{ANT05}. \\

\hd\ was observed serendipitously during all five \xmm\ pointings (see Table \ref{tab: obsx}). The source was recorded with the three {\sc e}uropean {\sc p}hoton {\sc i}maging {\sc c}amera ({\sc epic}) detectors \citep[PN, MOS1 and MOS2; for more details, see][]{STR01,TUR01}. The data were processed with the \xmm\ {\sc s}cience {\sc a}nalysis {\sc s}oftware ({\sc sas}, version 5.4.1). A check on proton flares resulted in the deletion of part of the exposure time. More details on the pipeline processing of these data can be found in \citet{RAA03}\footnote{In \citet{RAA03}, note that Column 4 of Table 5 actually gives the Julian dates in the format JD$-$2\,450\,000, not the MJDs.}.\\

\begin{table}
\centering
\caption{Journal of the \xmm\ observations of \hd. The first column gives the revolution during which the observations were taken, Column 2 the Julian Date (JD) at mid-exposure, Column 3 the phases (calculated from the average \he\ orbital solution, see Sect. 5), and the last three columns the exposure times in ks for each instrument. For more details, see \citet{RAA03}.
\label{tab: obsx} }
\begin{tabular}{l c c r r r}
\hline
Rev. & JD$-$2\,450\,000 & $\phi_{\rm avg}$& \multicolumn{3}{c}{Exp. Time (ks)}\\
& (days) &  & MOS1 & MOS2 & PN\\
\hline
115 & 1751.92 & 0.18 &  33.8&  30.6&  31.7\\
116 & 1753.56 & 0.37 &  11.2&   8.3&   9.4\\
283 & 2086.00 & 0.17 &  36.7&  36.7&  34.6\\
284 & 2089.05 & 0.53 &  42.1&  42.1&  39.6\\
285 & 2090.91 & 0.75 &  37.4&  37.4&  34.9\\
\hline
\end{tabular}
\end{table}

\section{The spectrum of \hd A}

Up to now, the visible spectrum of \hd\ was essentially studied through photographic observations. Following \citet{WAL73}, both stars of \hd\ have identical spectra and apparent magnitudes. Walborn further attributes an O6.5V((f)) type to each star. The following year, however, \citet{VRA} discovered that \hd A was cooler than \hd B. Unfortunately, no formal spectral type was assigned to \hd A by \citet{VRA}. \\ 

\begin{figure*}
\centering
\includegraphics[width=17cm]{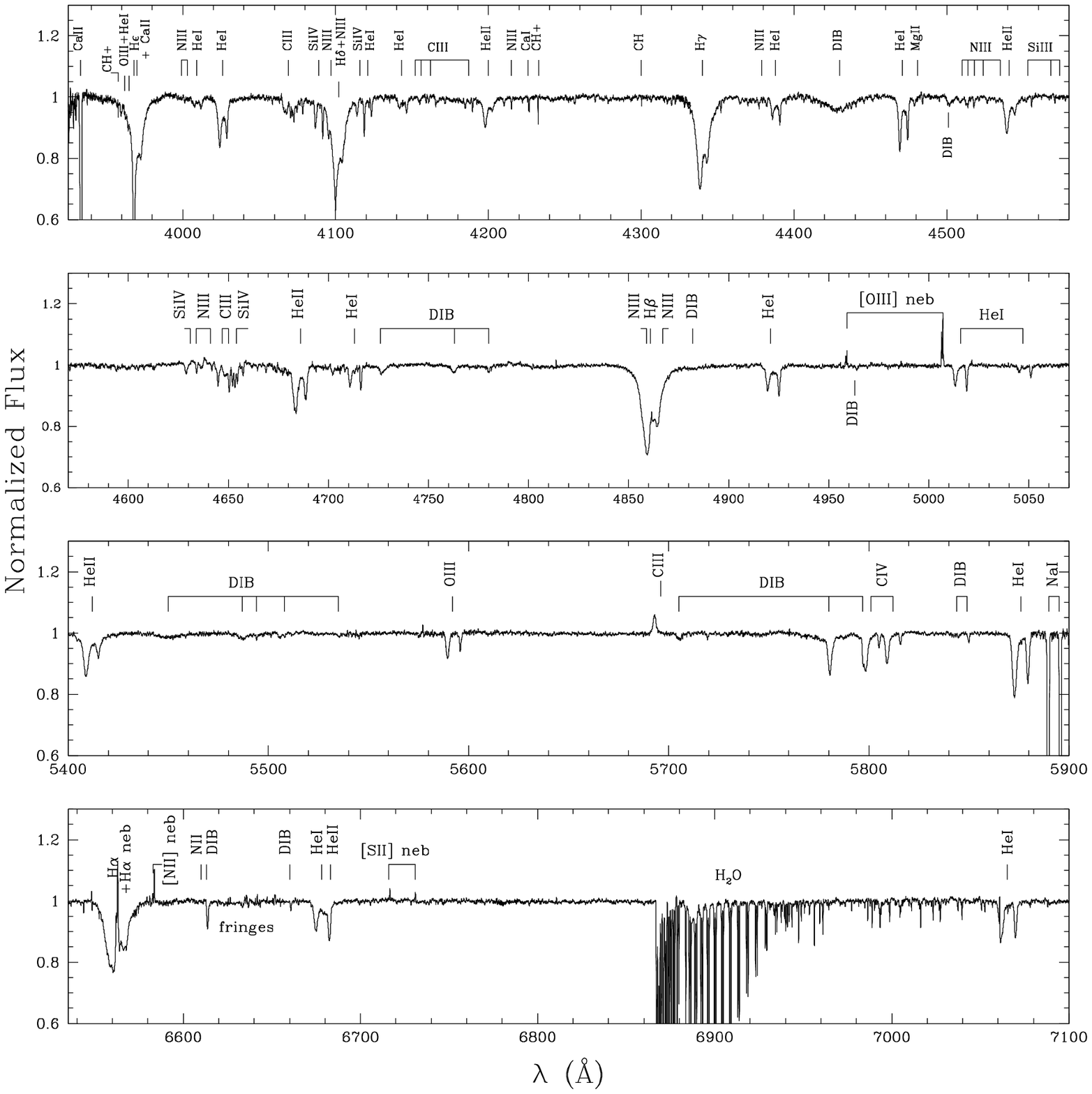}
\caption{\label{fig: speca} Spectrum of \hd A as observed with the FEROS spectrograph on HJD\,2\,453\,135.544 ($\phi \sim 0.7$). At this orbital phase, the spectrum of the most massive star is shifted towards the blue. The main lines are indicated at their rest wavelengths. 
}
\end{figure*}

We have investigated the spectrum of \hd A through high resolution echelle spectroscopy. Such a spectrum is shown in Fig. \ref{fig: speca}: the lines are clearly double, showing that this star is indeed a spectroscopic binary. Beside the lines of the Balmer series, the spectrum reveals many absorption lines of \he, \hee, \n, Si\,{\sc{iii}}, \si, C\,{\sc{iv}} and \o\ that are typical of O stars. It also reveals a number of diffuse interstellar bands (DIBs), as well as interstellar absorptions due to \he\,\l\,3889, Na\,{\sc{i}}, Ca\,{\sc ii}, CH and CH$^+$. There are only few emission lines in the spectrum of \hd A. Most of them are sharp nebular lines, associated with the Carina Nebula. We may note that the most massive component exhibits \ciii\,\l\,5696 in emission\footnote{The fact that only this star exhibits the \ciii\,\l\,5696 line in emission helps for the identification of the primary star. In addition, the presence of a second emission peak in this line clearly indicates pollution by \hd B, which also displays this line in its spectrum. It thus enables us to estimate the degree of contamination by \hd B in each spectrum of \hd A. }. Throughout this paper, we will refer to the more massive star as the primary, or \hd A1. \\

\begin{figure*}
\centering
\includegraphics[width=17cm]{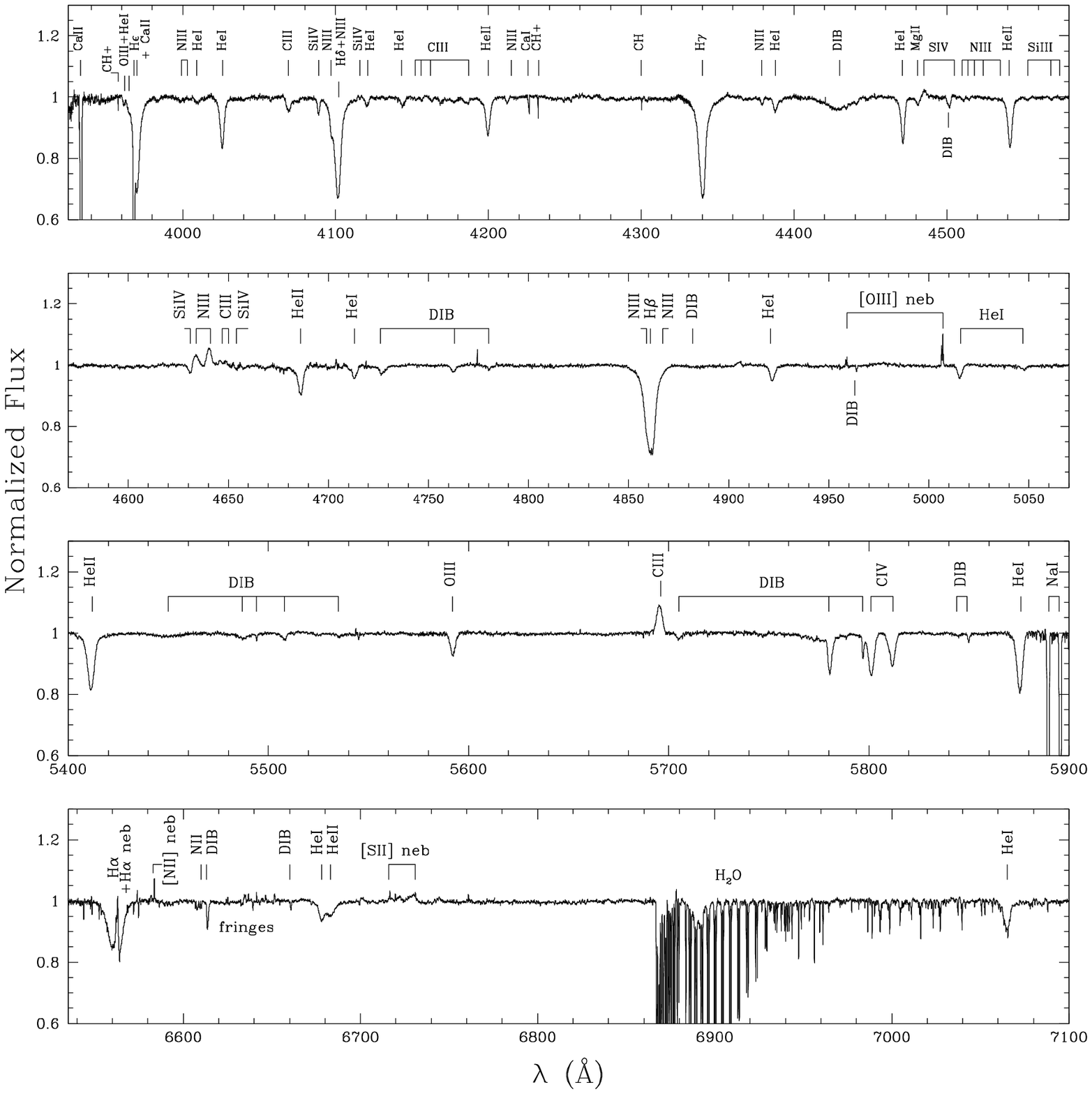}
\caption{\label{fig: specB} Spectrum of \hd B as observed with the FEROS spectrograph on HJD\,2\,453\,132.506. The same lines as in Fig. \ref{fig: speca} are identified.}
\end{figure*}

Only the least blended and higher quality spectra are suitable for the spectral classification. Following \citet{CA,CF} and \citet{MA1,MA2}, we adopt the $\log{[EW(4471)/EW(4542)]}$ classification criterion to determine the spectral types of the stars. For this purpose, we fitted these lines with a combination of several gaussians. This procedure enables to measure the radial velocities (RVs) rather securely, but the presence of small additional components (due to the contamination by the neighbouring \hd B) renders the determination of accurate equivalent widths (EWs) rather awkward. The measured ratios are thus rather dispersed, resulting in an approximate O7-7.5 type for the primary and an O7-8.5 type for the secondary at 1-$\sigma$ dispersion. To refine this classification further, we have compared their spectrum to the OB star atlas of \citet{WAL90}. For the primary star, \he\,\l\,4471 (resp. 4026) appears only slightly stronger than \hee\,\l\,4542 (resp. 4200). The reverse situation occurs for the \he\,\l\l\,4143, 4388 lines, which are weaker than the \hee\,\l\l\,4200, 4542 lines. The spectrum of \hd A1 is thus similar to that of HD\,48279 \citep{WAL90}, indicating an O8 type for this star. On the other hand, the secondary seems cooler: \he\,\l\,4471 (resp. 4026) is much stronger than \hee\,\l\,4542 (resp. 4200), while \he\,\l\,4143 is weaker than \he\,\l\,4388 but has a strength similar to those of \hee\,\l\l\,4200, 4542. \hd A2 thus resembles HD\,46202, an O9 star. These spectral types are slightly later than those derived from Conti's criterion. This difference is probably due to the contamination by the B component but we can not totally exclude that the stars might be one subtype earlier.\\

For the luminosity classification, we rely on the $\log{[EW({\rm Si\,IV}\,\lambda\,4089)/EW({\rm He\,I}\,\lambda\,4143)]}$ criterion introduced by \citet{CA}. For the same reason as before, the measured values of this ratio are rather scattered but they clearly indicate a main sequence classification. Note that the $\log{[EW({\rm He\,II}\,\lambda\,4686)]}+\log{[EW({\rm He\,I}\,\lambda\,4388)]}$ luminosity criterion introduced by \citet{MA1,MA2} was not used here: it relies on absolute EWs, which are difficult to determine because of the variable contamination by \hd B. The main sequence classification is further supported by the comparison of our spectra with the atlas of standards since, for both stars, \si\,\l\,4089 is quite faint whereas the \hee\,\l\,4686 line appears rather strongly in absorption. We thus conclude that \hd A is most probably an O8V+O9V binary. \\

\begin{table*}
\centering
\caption{Circular orbital solutions for \hd A as deduced from different lines. $K_1$ and $K_2$ yield the amplitude of the RV curves of the primary and secondary, respectively, whilst $\gamma_1$ and $\gamma_2$ correspond to the apparent systemic velocities of the two components. To account for a possible velocity gradient in the line formation region, we allowed $\gamma_1$ and $\gamma_2$ to be different. Quoted uncertainties are the 1-$\sigma$ error bars. $s_y/s_x$ gives the ratio of the secondary to primary uncertainties on the RVs.  
\label{tab: solorb} }
\begin{tabular}{l c c c c r r c}
\hline
Lines & $s_y/s_x$ & $P$   & $K_1$ & $K_2$ & \multicolumn{1}{c}{$\gamma_1$} &  \multicolumn{1}{c}{$\gamma_2$} & rms\\
      &           & (days)& (\kms)& (\kms)& (\kms)     &  (\kms)     &(\kms)\\
\hline
\he\,\l\,4026  &  0.8  & 8.5665 & 149.7$\pm$3.4 &  201.2$\pm$4.6  &   5.1$\pm$4.0  &  13.0$\pm$5.0 &14.8\\
\he\,\l\,4471  &  1.0  & 8.5662 & 151.5$\pm$2.4 &  202.2$\pm$3.2  &$-$1.4$\pm$2.5  &   2.2$\pm$3.1 &12.4\\
\hee\,\l\,4542 &  1.4  & 8.5664 & 150.8$\pm$4.8 &  198.1$\pm$6.4  &   4.7$\pm$7.7  &   2.9$\pm$9.6 &20.1\\
\hee\,\l\,4686 &  1.1  & 8.5663 & 150.8$\pm$2.4 &  201.1$\pm$3.2  &   9.1$\pm$2.5  &   3.0$\pm$3.0 &12.4\\
\he\,\l\,4921  &  0.8  & 8.5662 & 148.9$\pm$2.3 &  202.3$\pm$3.1  &$-$1.1$\pm$2.1  &   1.5$\pm$2.4 & 9.4\\
\he\,\l\,5016  &  0.8  & 8.5662 & 155.3$\pm$1.6 &  201.0$\pm$2.1  &   0.6$\pm$1.3  &$-$0.2$\pm$1.4 & 6.6\\
\hee\,\l\,5412 &  1.3  & 8.5666 & 150.8$\pm$4.1 &  192.9$\pm$5.2  &   9.0$\pm$7.5  &$-$0.2$\pm$9.5 &20.8\\
\o\,\l\,5592   &  1.2  & 8.5664 & 155.2$\pm$2.3 &  201.3$\pm$3.0  &   0.7$\pm$2.3  &$-$4.2$\pm$2.7 &10.4\\
\civ\,\l\,5801   &  1.2  & 8.5663 & 153.0$\pm$2.7 &  198.3$\pm$3.5  &$-$2.5$\pm$2.9  &$-$5.4$\pm$3.5 &11.3\\
\civ\,\l\,5812   &  1.4  & 8.5663 & 155.1$\pm$2.4 &  197.9$\pm$3.1  &$-$5.1$\pm$2.5  &$-$2.8$\pm$3.0 &10.0\\
\he\,\l\,5876  &  0.9  & 8.5663 & 152.5$\pm$2.4 &  202.2$\pm$3.2  &   3.9$\pm$2.4  &   3.1$\pm$2.8 & 9.8\\
\he\,\l\,7065  &  0.9  & 8.5663 & 155.8$\pm$2.2 &  201.4$\pm$2.9  &   2.9$\pm$2.0  &   4.1$\pm$2.2 & 8.4\\
\hline
\end{tabular}
\end{table*}

\section{The spectrum of \hd B}

A FEROS spectrum of \hd B is shown in Fig.\,\ref{fig: specB}. This spectrum is similar, but not identical to that of \hd A. \hd B actually exhibits a few more emission lines, e.g. the weak \s\,\l\,4485, \n\,\l\l\,4634, 4641 lines. As already remarked by \cite{VRA}, the star appears slightly hotter than its A neighbours, since the \he\,\l\,4471 line is weaker than the \hee\,\l\,4542 line.\\

Among the available spectra of \hd B, we again rely only on the high-quality FEROS data for the spectral classification. We used the same classification criteria as above: we determined $\log{[EW({\rm He\,I}\,\lambda\,4471)\,\,/\,\,EW({\rm He\,II}\,\lambda\,4542)]}\,\,\sim\,\,-0.14$, suggesting an O6.5 spectral type, while $\log{[EW({\rm Si\,IV}\,\lambda\,4089)\,\,/\,\,EW({\rm He\,I}\,\lambda\,4143)]}<0.1$, corresponding to a main sequence classification. We have also compared the spectrum of \hd B with those of the spectral atlas of OB standards \citep{WAL90}. In the spectrum of \hd B, the \hee\,\l\l\,4200, 4542 lines are much stronger than the \he\,\l\l\,4143, 4388 lines whereas \he\,\l\,4026 is stronger than \hee\,\l\,4200. This confirms the O6.5 type for this star. The relative weakness of the \si\,\l\,4089 line indicates a main sequence classification, again in agreement with Conti's criterion (which in principle does not apply to stars hotter than O7). Moreover, the \n\,\l\l\,4634, 4641 lines exhibit only a weak emission while the \hee\,\l\,4686 line is in absorption (with a lower intensity than that of the \hee\,\l\,4542 line), as is typically seen in some main sequence stars \citep{WAL90}. The presence of these latter lines require the addition of an ((f)) tag to the classification: \hd B is thus an O6.5V((f)) star, in agreement with the classification of \citet{WAL73}.  \\

The bottom part of Table \ref{tab: obs} gives the measured RVs of the \he\,\l\,4471 line on our 3 spectra of \hd B. While we find no obvious evidence for a companion in the spectrum, the observed RVs of all lines of \hd B are clearly variable. Except perhaps for the lower quality BME data, this is most likely not due to contamination by the A neighbour. Indeed, the contemporaneous spectra of the A star show two widely separated components, which are not (even slightly) detected in the B spectrum and could thus not hamper the RV determination. These RV variations might actually be caused by orbital motion in a binary. Unfortunately, the small number of B spectra does not enable us to further analyse the variability of this star, and especially to determine if these variations are periodic. Additional uncontaminated, high S/N data of \hd B would certainly allow to investigate the multiplicity of this star. \\

\section{Orbital solution for \hd A}

As mentioned before, we have determined the RVs of the lines by fitting them with a sum of gaussian profiles. In order to get a reliable orbital solution, we  considered 12 different lines: \he\,\l\l\,4026, 4471, 4921, 5016, 5876, 7065; \hee\,\l\l\,4542, 4686, 5412; \o\,\l\,5592 and \civ\,\l\l\,5801, 5812. We present in Table \ref{tab: obs} the RVs measured for the \he\,\l\,4471 line. Note that we have also measured the RVs of the most prominent interstellar lines: for our FEROS observations, the average RV of the CH$^+$\,\l\,4232 line is $5.3\pm0.4$~\kms, whereas for the B\&C data, the mean RV of the Ca~{\sc{ii}}\,\l\,3933 line is $-35\pm15$~\kms\ and for the BME data, the mean RV of the red component of the Ca~{\sc{ii}}\,\l\,3933 line is $-0.1\pm2.9$~\kms. We thus emphasize the very good temporal stability of the FEROS spectrograph, whose spectra constitute the majority of our sample.\\

Up to now, two orbital solutions for \hd\ have been published in the literature. \citet{LEV91} have calculated an SB1 solution for \hd, but without specifying to which component (either A or B) it refers. \citet{LUN03} failed to provide any improvement to the results of \citet{LEV91}\footnote{Note that in \citet{LUN03}, the figure showing the `new solution' for \hd\ is most probably erroneous. Moreover, in the same bulletin, the authors also provide an SB1 solution for Tr\,16-104 and claim to detect no RV variations for Tr\,16-110, although both stars were already known as triple systems prior to this publication \citep{RAU01,ALB02}. This may legitimately cast some doubt on the quality of their data.}. If these authors measured the RVs on combined spectra of \hd, as can be guessed from their description, it seems rather strange that they detected only one component (thus deriving an SB1 solution) since the presence of multiple components is easily spotted even in our medium resolution B\&C spectra. In the rest of this paper, we will not discuss these analyses further.\\

\begin{table}
\centering
\caption{Orbital and physical parameters of \hd A as deduced from the mean RVs for a circular orbit. The usual notations are used, $R_{RL}$ being the Roche lobe radii of the stars and $T_0$ corresponding to the conjunction with the primary star in front. The error on the period is calculated from the observational time base (see \S 5.1) and all other errors are estimated 1-$\sigma$ deviations calculated for a fixed $s_y/s_x$ ratio and a fixed period (the ones that give the lowest residuals). Note that the apparent systemic velocities ($\gamma_1$ and $\gamma_2$) of each line were subtracted prior to averaging. \label{tab: solorbfin} }
\begin{tabular}{l c }
\hline
$P$                        & 8.5663$\pm$0.0040~days\\
$T_0$ (HJD                 & 3001.098\\
$-$2\,450\,000)            & $\pm$0.011\\
$s_y/s_x$                  & 0.9\\
$m_1/m_2$                  & 1.31$\pm$0.02\\
$K_1$                      & 152.9$\pm$1.7~\kms\\
$K_2$                      & 200.6$\pm$2.2~\kms\\
$\gamma_1$ (after sub.)    & 0.02$\pm$1.4~\kms\\
$\gamma_2$ (after sub.)    & 0.05$\pm$1.6~\kms\\
$a_1 \sin i$               & 25.9$\pm$0.2~R$_{\odot}$\\
$a_2 \sin i$               & 33.9$\pm$0.4~R$_{\odot}$\\
$m_1 \sin^3 i$             & 22.2$\pm$0.6~M$_{\odot}$\\
$m_2 \sin^3 i$             & 17.0$\pm$0.4~M$_{\odot}$\\
\vspace*{-0.25cm}&\\
$\frac{R_{RL}^1}{a_1+a_2}$ & 0.4026$\pm$0.0015\\
\vspace*{-0.25cm}&\\
$\frac{R_{RL}^2}{a_1+a_2}$ & 0.3557$\pm$0.0014\\
\vspace*{-0.25cm}&\\
rms &7.7~\kms \\
\hline
\end{tabular}
\end{table}

\subsection{Period determination}

Once the RVs are measured, the next step is to determine the period of the orbital motion. To this aim, we have applied the method of \citet[ hereafter LK]{LK} as well as a Fourier-type analysis (method of \citealt{HMM}, hereafter HMM, see also the comments by \citealt{GOS}). Since confusion could be a possible issue, these period search algorithms were first applied to the time series of the absolute RV differences $|RV_1-RV_2|$: for each of the lines quoted above, a period of about 4.3~days was then detected. Using this preliminary period and the RVs of the \ciii\,\l\,5696 line, we then identified the primary and secondary, and applied the period search on the actual differences $RV_1-RV_2$. Both the Fourier analysis and the LK algorithm agree on a value of $P=8.565$~days for all lines. \\

Our dataset consists of 42 spectra spread over 2600~days for the \he\,\l\,4471 and \hee\,\l\,4686 lines (where there exist B\&C data in addition to the FEROS and BME observations) and of 36 spectra spread over 1836~days for the other lines. These observational time bases correspond respectively to natural peak widths $\Delta\nu$ of 0.00038~d$^{-1}$ and 0.00054~d$^{-1}$. The full width at half maximum of the actual peaks in the periodogram is in agreement with these theoretical values. We adopt an uncertainty of one tenth of the peak's width, i.e. $P = 8.565\pm0.003$ or 0.004~days, the uncertainty depending on whether or not the B\&C data are considered. \\

\subsection{Orbital elements}

To derive the orbital elements, we used a modified version of the \citet{WOL67} method. The algorithm, presented by \citet{SHR03}, allows a  weighting that differs for the secondary and primary RVs. In the present paper, we have adopted a period value and a relative (primary vs secondary) weight that yield the lower residuals. In addition, we accounted for the quality of the measurements by  attributing lower weights to the observations where the components are blended (i.e. only one measurement is available for both stars), noisy (e.g. BME and B\&C data), or heavily contaminated by \hd B. \\

\begin{figure}
\centering
\includegraphics[width=0.43\textwidth]{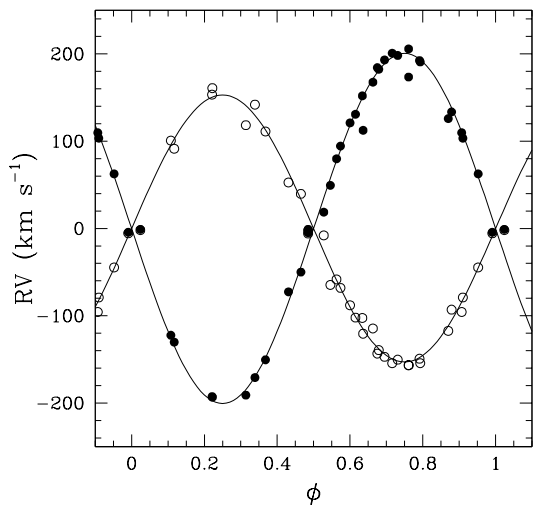}
\caption{\label{fig: bestfit} Radial velocity curves corresponding to the averaged RVs of \hd A and assuming $e = 0.0$ (see Table\,\ref{tab: solorbfin}). The open and filled circles correspond to the observed mean RVs of the primary and secondary, respectively. Note that we used a reference frame with zero systemic velocity (see text).}
\end{figure}

At first, we fitted the RVs from each line with a general, eccentric solution. However, the computed eccentricity $e$ turned out to be very small ($< 0.03$), with a null eccentricity within less than two sigma. Therefore, a circular orbit was assumed for \hd A. The results of these non-eccentric fits are displayed in Table \ref{tab: solorb} for each line. They are compatible with each other, within the errors.\\

Finally, we tried to get more accurate results by averaging the RVs from different lines. This was done after subtracting the apparent systemic velocities derived for each line from the individual fits. Note that the lines yielding the largest residuals (i.e. \hee\,\l\l\,4542, 5412 and \he\,\l\,4026) were not considered. The lower resolution B\&C data were also discarded in order to have a more homogeneous set of RVs. Averaging all other observations allows to obtain an rms residual of 8~\kms. This best solution, including a reappraisal of the period ($P_{\rm final}=8.5663\pm0.0040$~days), is presented in Table \ref{tab: solorbfin} and is shown in Fig. \ref{fig: bestfit}; it will be adopted throughout the rest of the paper. \\

The minimum masses $m\,\sin^3 i$ derived for the primary and secondary are 22.2 M$_{\odot}$ and 17.0 M$_{\odot}$, respectively (see Table \ref{tab: solorbfin}). Comparing these values to typical masses of O8V and O9V stars allows to estimate the inclination of the system. However, one must be careful when doing so, since the analysis of the binaries in the Tr\,16 cluster (HD\,93205, Paper II and Tr\,16-104, Paper III) has shown that the typical masses presented by well-known references like e.g. \citet{HP89} are generally too large compared to the masses derived from the study of binary systems. We therefore decide to adopt the values determined for detached eclipsing systems \citep{gies}: $\sim$22~M$_{\odot}$ for O8V stars and $\sim$18~M$_{\odot}$ for O9V stars. Such values are very close to the minimum values found for \hd A 1 and 2, indicating a large inclination, most probably $>75^{\circ}$. With such an inclination, the system should undergo eclipses. Unfortunately, no photometric study of \hd\ has been conducted up to now: additional observations are thus needed. They will help to constrain further the physical parameters of \hd A.\\


\section{\xmm\ observations of \hd}

\subsection{Spectral analysis}

\begin{figure}
\centering
\includegraphics[width=0.43\textwidth]{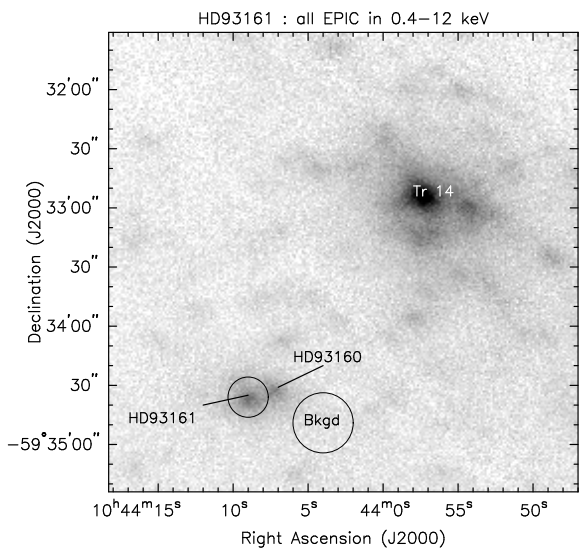}
\caption{\label{fig: imx} Combined {\sc epic} image showing \hd, HD\,93160 and their neighbourhood. The bright X-ray source to the top right is the core of the Tr\,14 cluster, that contains notably Tr\,14-1 and HD\,93129AB. The source and background regions used to extract the X-ray spectrum of \hd\ are also shown. Note that this image was exposure-corrected.}
\end{figure}

Due to the proximity of HD\,93160, the spectrum of \hd\ was extracted from a circular region with a radius limited to 10\arcsec, as is shown in Fig. \ref{fig: imx}. Due to a poor calibration of the Point Spread Function of the {\sc epic} instruments (especially PN) at large off-axis angles, we extracted MOS and PN spectra in the extended source mode, i.e. without any EEF (encircled energy fraction) correction. For the chosen region, we estimate that this correction amounts to a factor $\sim1.8$. The background spectrum was obtained by selecting events from a circular region spatially offset from the source (see Fig. \ref{fig: imx}). \\

\begin{table*}
\centering
\begin{minipage}{140mm}
\vspace{5mm}
\caption{Best-fitting models and X-ray fluxes at Earth for each \xmm\ observation of \hd. The fitted model has the form {\tt wabs($N_{\rm int}^{\rm H}$)*(mekal(k$T_1$)+wabs($N^{\rm H}$)*mekal(k$T_2$))}, with {\tt wabs($N_{\rm int}^{\rm H}$)}$=4.5\times10^{21}$~cm$^{-2}$ \citep{DIP94}. Quoted fluxes are in the 0.4$-$10.~keV energy range and are not corrected for EEF (see Sect. 2.2). The unabsorbed fluxes $f_X^{\rm unabs}$ are dereddened only considering the interstellar absorbing column. The quoted errors corresponds to 1-$\sigma$ deviations.
\label{tab: table_fit}}
\setlength{\tabcolsep}{1.5mm}
\begin{center}
\begin{tabular}{ccccclcc}
\hline
Rev. & $\phi_{\rm avg}$ & k$T_1$         & $N_2^{\rm H}$  & k$T_2$         & $\chi^2_{\rm \nu}$(dof) & $f_X^{\rm abs}$ & $f_X^{\rm unabs}$\\
     & & keV            & $10^{22}$~cm$^{-2}$ & keV            &                               & \multicolumn{2}{c}{($10^{-13}$~erg\,cm$^{-2}$\,s$^{-1}$)}  \\
\hline
 115 & 0.17 &$0.26\pm 0.02$ & $0.43\pm 0.13$ & $0.71\pm 0.11$ & 1.14 (132) & 0.85 & 4.91 \\
 116 & 0.37 &$0.29\pm 0.03$ & $0.44\pm 0.66$ & $0.93\pm 0.67$ & 1.26 (37)  & 0.77 & 4.46 \\
 283 & 0.17 &$0.27\pm 0.03$ & $0.45\pm 0.14$ & $0.67\pm 0.08$ & 0.87 (106) & 0.92 & 4.67 \\
 284 & 0.53 &$0.29\pm 0.02$ & $0.44\pm 0.13$ & $0.71\pm 0.12$ & 0.89 (122) & 0.96 & 4.72 \\
 285 & 0.73 &$0.29\pm 0.01$ & $0.86\pm 0.16$ & $0.79\pm 0.11$ & 1.18 (139) & 0.89 & 4.57 \\
\hline
\end{tabular}
\end{center}
\end{minipage}
\end{table*}

\begin{figure}
\centering
\includegraphics[bb=112 40 552 700,clip, width=0.3\textwidth, angle=270]{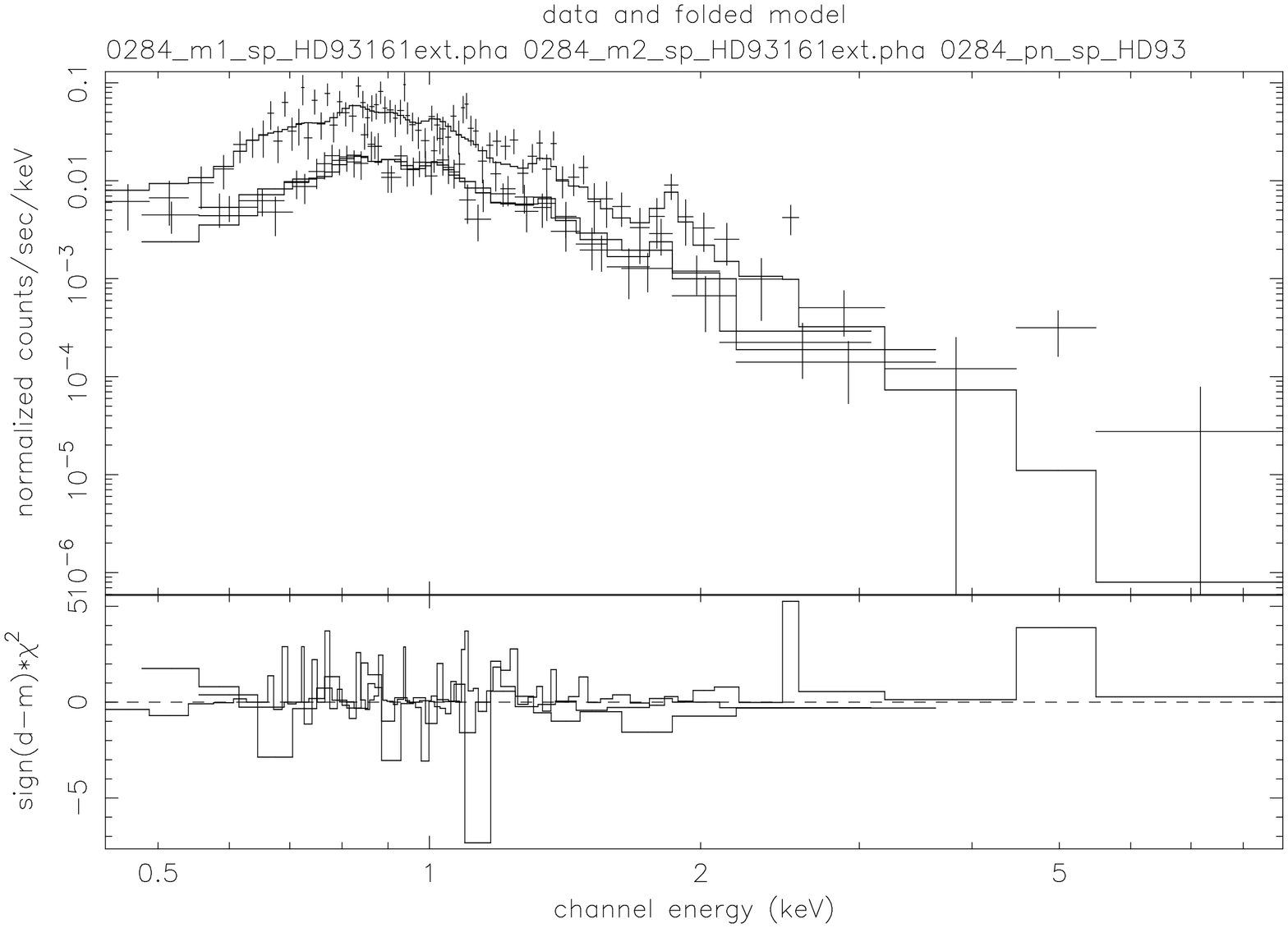}
\caption{{\sc epic} spectra of \hd\ obtained for Rev.\,\#284. In the top panel, the upper and lower data correspond to the PN and MOS spectra, respectively, and the solid lines yield the best-fitting absorbed two-temperature model (see Table \ref{tab: table_fit}). The bottom panel shows the contributions of individual energy bins to the $\chi^2$ of the fit. The contributions are carried over with the sign of the deviation (in the sense data minus model).}
\label{fig: specx}
\end{figure}

The X-ray data were then analysed with the {\sc xspec} software (version 11.2.0). For each observation, the {\sc epic} spectra were fitted simultaneously by a two-temperature thermal plasma {\sc mekal} model \citep{MEW85,KAA82} of the form {\tt wabs$_{\rm int}$*(mekal$_1$+wabs*mekal$_2$)} where {\tt wabs$_{\rm int}$} was fixed to the interstellar absorption determined by \citet[ $N_{\rm int}^{\rm H}=4.5\times10^{21}$~cm$^{-2}$]{DIP94}. The second {\tt wabs} model is included to account for the possibility of additional absorption for the hard component of the spectrum. The abundances were set to solar. The best-fitting model parameters are presented in Table \ref{tab: table_fit} and the spectra from Rev.\,\#284 are shown in Fig. \ref{fig: specx} along with their best-fitting model. The spectral parameters and the fluxes are fairly stable throughout the five datasets. The parameters are quite typical for a single O-type star, with k$T_1\sim0.3$~keV along with a slightly higher temperature component. Our results are in agreement with the preliminary conclusions of \citet{ALB03} concerning the first two datasets. \\

\subsection{Variability in the X-ray domain}

We also investigated the variations of the X-ray emission of \hd, first on a short time scale (within each pointing) and then on the long-term range (between the different pointings). For each dataset, the source and background events were extracted using the same regions as those defined for the spectrum analysis. After taking into account the flaring intervals (see Sect.\,2.2), these event lists were used to construct background-subtracted lightcurves that were analysed by various methods \citep{SAN04}. No significant short-term variability was consistently detected in the three instruments. 

As a next step, we searched for variations between the different pointings. To this aim, we compared the mean count rates as well as the fluxes determined by spectral fitting (see Table \ref{tab: table_fit}): no significant long-term changes from one pointing to another were detected.

\subsection{Wind interactions in \hd ?}

We found no clear evidence that the components of \hd\ are interacting in any manner. The visible spectrum presents only few emission lines and, in particular, no \ha\ emission is seen that might arise in a wind collision zone \citep{THA97,SAN01}. Moreover, the minimum Roche lobe radii, $R_{RL}\,\sin i$, determined for the primary and secondary of \hd A are 24.1$\pm$0.2 and 21.3$\pm$0.2, respectively (see Table \ref{tab: solorbfin}). These large values contrast with the small typical radii ($<$10~R$_{\odot}$) of O8-9V stars \citep{HP89,MOR01,RAU01,gies,SAN05b}, indicating that this object is a well detached binary system without any ongoing Roche lobe overflow. In addition, \hd\ is not a particularly bright X-ray source, since its unabsorbed luminosity amounts only to $\sim6.3\times10^{32}$~erg~s$^{-1}$ (after EEF correction and for a distance of 2.5~kpc, see next section), or $\log(L_X/L_{\rm bol})=-6.6$ (using the $L_{\rm bol}$ determined in the next section). 

We can compare this observed X-ray luminosity to the emission expected for a wind collision in \hd A. First, we determine the mass-loss rates $\dot M$ and terminal velocities $v_{\infty}$ using the recipes of \citet{VI00,VI01} and the probable radii, masses, temperatures, and bolometric luminosities of the stars (see Sects. 5.2 and 7). Using the formalism of \citet{PIT02}, we then find that, in the case of \hd A, a radiative wind collision would lead to an X-ray emission with $L_X\sim2\times10^{34}$~erg~s$^{-1}$. However, the physical parameters of \hd A (strengths of the winds, separation between the stars) should rather lead to a slightly adiabatic shock: the X-ray emission associated with such a collision could thus be significantly lower than in the radiative case. In fact, the total luminosity derived after applying the `canonical' $L_X$--$L_{\rm bol}$ relation \citep[see][ and \citealt{SAN05}]{BER96} to the three massive stars of \hd\ is $\sim3.4\times10^{32}$~erg~s$^{-1}$, indicating that the observed X-ray emission does not display a large excess attributable to a wind interaction. This conclusion is further supported by the lack of hard X-ray photons and the constancy of the X-ray emission. We might note however that our X-ray observations do not sample the whole orbital cycle: as was the case for CPD$-41^{\circ}7742$ \citep{SAN05b}, the signature of a wind collision could be limited to a few crucial phases. \\

Finally, \citet{BEN04} observed the Carina nebula with ATCA at 3 and 6\,cm aiming at the detection of non-thermal radio emission but \hd\ was not detected in this survey, contrary to what would be expected if there was a wind collision between \hd A and \hd B. Therefore, from the available data, we conclude that there is probably no strong wind interaction in \hd. \\

\section{Fundamental parameters}

Using the temperature scale of \citet{MAR02}, we can convert the spectral types derived above into effective temperatures of $38200 \pm 700$\,K, $35250 \pm 700$\,K and $33700 \pm 800$\,K for \hd B, \hd A1 and \hd A2 respectively. The quoted uncertainties correspond to half a subclass for the components of \hd A and to the range of temperatures for the O6.5\,V class according to Martins et al. 

In order to derive the luminosities, we need the observed magnitude, the reddening, the brightness ratios between the different components, the bolometric corrections and an estimate of the distance. The WEBDA\footnote{available at the url http://obswww.unige.ch/webda/} \citep{MER88,MER92} database lists the results of several photometric studies that provide the apparent $V$ magnitude of \hd\ \citep{WAL73,FEI73,KAL93,FOR78,MAS93}. On average, we find $V = 7.84 \pm 0.02$ for the entire \hd\ system. On the other hand, \citet{VAZ96} quote $V = 8.56$ and $V = 8.60$ for \hd A and \hd B respectively. These magnitudes yield $V = 7.83$ for the entire system (components A + B) in good agreement with the mean value hereabove. \citet{MAS98} found $V = 9.0$ for \hd A and $V = 9.1$ for \hd B. If these values were correct, the total magnitude of \hd\ should be 8.3 in contradiction with the other observations. In the following, we therefore adopt apparent magnitudes of $8.56 \pm 0.02$ and $8.60 \pm 0.02$ for \hd A and B respectively.

The $(B - V)$ colour indices of \hd A and \hd B were found to be 0.20 and 0.23 respectively \citep{VAZ96}. Comparing with typical colour indices of O stars \citep{SK82}, we hence adopt $E(B - V) = 0.52$ and $E(B - V) = 0.55$ for \hd A and \hd B. 

To estimate the relative brightness of the components of \hd A in the visible spectral range, we evaluate the ratio of the EWs of the primary and secondary components for a number of lines in our 2004 FEROS spectra (the least contaminated by the B component). We use the \he\,\l\l\,4471, 5876, \hee\,\l\,4542 and \o\,\l\,5592 lines and we compare the EW ratios to the values for typical EWs evaluated from the compilation provided by \citet{CA} and \citet{CON74}. In this way, we estimate a brightness ratio of $1.7 \pm 0.3$ between the primary and secondary of \hd A. This results in apparent magnitudes of $9.06 \pm 0.07$ and $9.64 \pm 0.12$ for the primary and secondary components respectively.

The distances to the clusters Trumpler\,14, Trumpler\,16, Collinder\,228 and Collinder\,232 are still poorly known. How controversial the issue of the distances and even the actual existence of these clusters is, can be illustrated by a brief overview of the literature. \citet{WAL95} summarised the situation as follows: Tr\,14 is younger than Tr\,16, but both clusters lie at the same distance; neither Collinder\,228 nor 232 are genuine clusters, the former is part of Tr\,16, whereas the latter contains stars of Tr\,14 and Tr\,16. The ambiguities on the distances of Tr\,14 and Tr\,16 are at least partly due to the uncertainties on the extinction towards these clusters. Using {\it UBV} photometry, \citet{MAS93} inferred a distance of 3.2\,kpc for Tr\,14 and Tr\,16. Assuming a normal extinction law and based on Str\"omgren and H$\beta$ photometry, \citet{KAL93} derived distances of individual stars in Tr\,16. These authors reached a radically different conclusion: since our line of sight towards Tr\,16 is almost parallel to the molecular cloud ridge in the Carina spiral arm, the stars might not belong to an actual cluster but could instead lie along an extended star formation region which is projected on a small area on the sky. Recently, \citet{CAR04} presented a study of Tr\,14, Tr\,16 and Collinder\,232 based on $UBVRI$ photometry. These authors argued that a unique reddening law is not appropriate to study the whole region and they estimated different values of the selective extinction $R_V = A_V/E(B - V)$ for each cluster. By comparing their observations with empirical main-sequences, they derived distances of about 2.5\,kpc for Tr\,14 and Collinder\,232, but a much larger distance of about 4\,kpc for Tr\,16. The latter result is clearly at odds with previous distance determinations. In particular, the studies of two eclipsing binaries in Tr\,16 \citep[Tr\,16-1,][ and Tr\,16-104, \citealt{RAU01}]{FRE01} allowed to obtain absolute luminosities of these stars. Adopting bolometric corrections from \citet{HUM84} and assuming $R_V = 3.3$, a distance of about 2.5\,kpc is inferred for these binaries, in excellent agreement with the Tr\,16 distance found by \citet{TAP03}. In the following, we shall therefore adopt a distance of 2.5\,kpc for \hd. 

For Collinder\,232, \citet{CAR04} infer $R_V = 3.73 \pm 0.03$, whereas they estimate $R_V = 3.83 \pm 0.33$ for \hd\ and $R_V = 3.60 \pm 0.27$ for its close neighbour on the sky HD\,93160. Adopting an average value of $R_V = 3.72 \pm 0.16$, we obtain absolute magnitudes of $-4.86 \pm 0.11$, $-4.29 \pm 0.15$ and $-5.44 \pm 0.09$ for \hd A1, A2 and B respectively. Finally, adopting the bolometric corrections from \citet{HUM84}, we obtain bolometric luminosities of $\log{L_{\rm bol}^{A1}/L_{\odot}} = 5.25 \pm 0.04$, $\log{L_{\rm bol}^{A2}/L_{\odot}} = 4.93 \pm 0.06$ and $\log{L_{\rm bol}^B/L_{\odot}} = 5.53\pm 0.04$.

\begin{figure}
\centering
\includegraphics[width=0.43\textwidth]{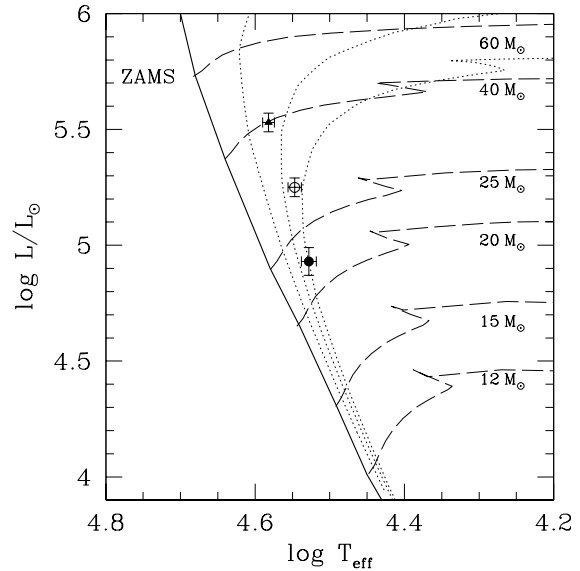}
\caption{Location of components A1 (open circle), A2 (filled circle) and B (filled triangle) of \hd\ in the Hertzsprung-Russell diagram. The evolutionary tracks are from \citet{SCH92}. Isochrones corresponding to ages of 2, 3 and 4 Myr (dotted lines) are also drawed. \label{fig: hrd}}
\end{figure}

Using these values, we have plotted in Fig.\,\ref{fig: hrd}, the components of \hd\ in the Hertzsprung-Russell diagram along with the evolutionary tracks from \citet{SCH92} for solar metallicity. This diagram suggests that the stars have an age of 3-4~Myr and are thus slightly evolved off the zero age main sequence (ZAMS), unlike other stars of the Tr\,16 cluster, e.g. the components of Tr\,16-104 which were found to lie very close to the ZAMS \citep{RAU01}. A crude interpolation between the evolutionary tracks yields evolutionary masses of 30.1$\pm$1.1~M$_{\odot}$ and 22.4$\pm$0.8~M$_{\odot}$ for the primary and secondary, respectively. This corresponds to a mass ratio of 1.34$\pm$0.07, in excellent agreement with the observed ratio of 1.31$\pm$0.02. The orbital inclination corresponding to these evolutionary masses would be 65$\pm$2$^{\circ}$, and this lower value of the inclination would prohibit eclipses. A photometric study of \hd\ is thus necessary in order to assess the masses but we might already note that in the case of the binary HD\,93205 (Paper II), evolutionary masses are larger than the true masses. Finally, we caution that the luminosity of \hd B might be overestimated if the star turned out to be another binary system.

\section{Conclusions}

We have presented a thorough spectroscopic investigation of \hd, based on medium  and high resolution data obtained over a period of 7 years. Although the small separation between the two components of \hd\ did not enable us to get totally uncontaminated spectra, we have shown that \hd\ actually harbours a detached SB2 system (\hd A) and a star displaying RV variations (\hd B). \\
The primary (resp. secondary) star of \hd A is most probably of spectral type O8V (resp. O9V). Using the RVs measured for each component of \hd A, we provide the first orbital solution for this system. The orbital period is 8.566$\pm$0.004~days and the minimum masses for the primary and secondary stars are 22.2$\pm$0.6~M$_{\odot}$ and 17.0$\pm$0.4~M$_{\odot}$, respectively. Such large minimum values, compared to typical masses for O8V and O9V stars, suggest a high inclination of the orbit. It is therefore possible that \hd A undergoes eclipses. A follow-up study should be undertaken to check the photometric behaviour of \hd A. \\
The visual companion of \hd A, \hd B, is of type O6.5V((f)) and is thus slightly hotter than its neighbours. \hd B presents radial velocity variations, but no fully obvious signature of a companion could be found in its spectrum. The small number of our observations of this star did not enable us to investigate the periodicity of the velocity changes. Additional observations are required to check the binary status of \hd B. \\
Considering that \hd\ lies at a distance of 2.5~kpc, we found that all three components of \hd\ are slightly evolved off the ZAMS. In the X-ray domain, \hd\ appears as a moderately bright X-ray source displaying a two-temperature spectrum and no significant variability, suggesting an X-ray emission predominantly intrinsic to the stellar winds of the individual components. \\

\section*{Acknowledgments}

The Li\`ege team acknowledges support from the Fonds National de la Recherche Scientifique (Belgium) and through the {\sc PRODEX XMM-OM} and Integral projects. This research is also partly supported by contracts P5/36 ``P\^ole d'Attraction Interuniversitaire'' (Belgian Federal Science Policy Office). IIA acknowledges support from the Russian Foundation for Basic Research (project No 02-02-17524) and the Russian LSS (project No Nsh-388.2003.2).\\

\label{lastpage}

\end{document}